\documentclass{llncs}

\usepackage[utf8]{inputenc}
\usepackage[T1]{fontenc}

\usepackage[english]{babel}

\usepackage{todonotes}

\usepackage{graphicx}
\usepackage{url}

\usepackage[babel]{csquotes}

\usepackage{amsmath}
\usepackage{amsfonts}
\usepackage{amssymb}
\usepackage{pifont}

\usepackage{listings}
\AtBeginDocument{}
\usepackage{hyperref}
\usepackage[capitalise,noabbrev]{cleveref}

\usepackage{xspace}
\usepackage{booktabs}

\usepackage{breakurl} 

\newcommand{\cmark}{\ding{51}}
\newcommand{\xmark}{\ding{55}}
\newcommand{\partialmark}{(\cmark)\xspace}

\newcommand{\ignore}[1]{} 

\newcommand{\constring}{\textsc{ConString}}

\begin{document}

\title{Towards Constraint Logic Programming over Strings for Test Data Generation}
\author{Sebastian Krings\inst{1} \and Joshua Schmidt\inst{2} \and Patrick Skowronek\inst{3} \and Jannik Dunkelau\inst{2} \and Dierk Ehmke\inst{3}}
\institute{
Niederrhein University of Applied Sciences\\
M\"onchengladbach, Germany\\
\email{sebastian.krings@hs-niederrhein.de}
\and
Institut f\"ur Informatik, Heinrich-Heine-Universit\"at\\
D\"usseldorf, Germany
\and
periplus instruments GmbH \& Co. KG\\
Darmstadt, Germany
}

\maketitle

\begin{abstract}
In order to properly test software, test data of a certain quality is needed.
However, useful test data is often unavailable:
Existing or hand-crafted data might not be diverse enough to enable desired test cases.
Furthermore, using production data might be prohibited due to security or privacy concerns or other regulations.
At the same time, existing tools for test data generation are often limited.

In this paper, we evaluate to what extent constraint logic programming can be used to generate test data, focussing on strings in particular.
To do so, we introduce a prototypical CLP solver over string constraints.
As case studies, we use it to generate IBAN numbers and calender dates.
\end{abstract}

\section{Introduction}\label{introduction}
Gaining test data for software tests is notoriously hard.
Typical limitations include lack of properly formulated requirements or the
combinatorial blowup causing an impractically large amount of test cases needed to cover the system under test (SUT). 
When testing applications such as data warehouses, difficulties stem from the amount and quality of test data available and the volume of data needed for realistic testing scenarios~\cite{ElGamal:2013:DWT:2457317.2457319}.
Artificial test data might not be diverse enough to enable desired test cases~\cite{Haftmann2007},
whereas the use of real data might be prohibited due to security or privacy concerns or other regulations~\cite{4086351}, e.g, the ISO/IEC 27001~\cite{ISO27001}. Further challenges have been identified by Khan and ElMadi~\cite{testingthesis}.

In consequence,
to properly test applications one often has to resort to
artificial test data generation~\cite{4086351}.
However, existing tools are limited as they
\begin{itemize}
  \item generate data that does not cover the desired scenarios~\cite{Haftmann2007},
  \item are specialized and lack options for configuration and adaptation~\cite{Houkjr2006SimpleAR}, and
  \item generate an amount of data that is unrealistic for the SUT~\cite{dataamount}.
\end{itemize}

In this paper, we evaluate to what extent constraint logic programming could be used for test data generation, in particular for generating strings.
We are not concerned with software testing itself.

\section{Test Data}%
\label{sec:testdata}
The International Software Testing Qualifications Board (ISTQB) describes test data as data created or selected to satisfy the preconditions and inputs to execute one or more test cases~\cite{SpillnerLinz05}. 
Test data may belong to the following categories: 
\begin{itemize}
  \item status data, files or surrounding systems required for a reusable start state,
  \item input data transferred to a test object during test execution,
  \item output data returned by a test object after execution,
  \item production data, which is deducted from the production system.
\end{itemize}

Production data is often used for testing 
as it provides obvious test cases and can be gathered easily. 
However, using production data does not lead to thorough testing, e.g.,
it never contains dates in the future.
While production data can be anonymized, it is hard to guarantee that de-anonymization is impossible.
Furthermore, production data may be biased.

Those problems can be solved by generating synthetic data.
The implementation of a test data generator for each specific problem is cumbersome. 
One just wants to describe the problem at hand 
without implementing the actual data generation. 
We therefore consider constraint programming to be appropriate 
for implementing general test data generators.
However, generating synthetic data remains a complex task 
as it involves thoroughly specifying constraints the data needs to fulfill 
in order to derive high quality test data.

\subsection{Test Data Generators}
The generation of synthetic test data can be supported by different test data generators~\cite{SpillnerLinz05}:
Database-based generators synthesize data according to database schemata or create partial copies of database contents,
i.e., they rely on production data.
Interface-based generators analyze the test object's API and determine the definition areas of input parameters to derive test data from.
Again, test oracles cannot be derived.

Code-based generators take the source code of the SUT into account, which has disadvantages.
For instance, it prevents oracle generation and is unable to work with
source code that is not available (e.g., for foreign libraries).
Furthermore, code-based generators are a weak test base, especially lacking the intellectual redundancy necessary for testing (four-eyes principle)~\cite{modelPretschner}, i.e., the understanding of how a system is supposed to work and how it is implemented are necessarily identical if tests are generated purely based on code.

Specification-based generators generate test data and oracles based on specifications written in a formal notation.
A specification based generator could thus generate data that replaces production data.
The quality of the test data is ensured by the model and the correctness of the solver. This includes quality aspects such as conformity and accuracy. 
To build such a generator, constraint solving over all needed data types is required. 

\subsection{Requirements Towards Solvers}\label{requirements}
To gain a sensible set of requirements for a string constraint solver for test data generation, we decided to look at the feature set of Oracle SQL.
The reasoning behind this is as follows:
SQL was designed for the description of complex data flows and is therefore suited as a modeling language for test data generation~\cite{Testdatenmanagement}.
It is widely used by developers, test data specialists and technical testers, i.e., they would be able to use it as a possible input language for generation tools.
Additionally, SQL statements can easily be extracted from source code
and can thus be used to automatically generate test data for given applications.
Furthermore, SQL is declarative and offers a good level of abstraction.

There are several types of strings in Oracle SQL%
\footnote{\url{https://docs.oracle.com/en/database/oracle/oracle-database/19/cncpt/tables-and-table-clusters.html\#GUID-A8F3420D-093C-449F-87E4-6C3DDFA8BCFF}},
in particular, unbounded unicode strings.
In addition, other data types are required for practical test data:
integers, fixed point numbers, reals and dates.
There are no booleans in SQL, however, booleans ease encoding complex SQL conditions into constraints.
\footnote{For now, we are not interested in supporting binary large objects (e.g., images stored inside the database), since
SQL does not include operations on them and their semantics are usually invisible to the applications.}
Oracle SQL lists 54 functions on strings%
\footnote{\url{https://docs.oracle.com/en/database/oracle/oracle-database/19/sqlrf/Functions.html\#GUID-D079EFD3-C683-441F-977E-2C9503089982}}.
 The ones we are interested in are
 \begin{itemize}
   \item CONCAT: concatenation of strings
   \item LENGTH: returns the length of a string
   \item REGEXP: tests, whether a string matches a given regular expression or not
   \item SUBSTRING: returns a substring with given start position and length
   \item TO\_NUMBER: convert a string to number and vice versa
 \end{itemize}  

Other operations can often be implemented with these functions or are not of interest for test data generation.
REGEXP requires the solver to process regular expressions. 
The constraint handlers for all types must interwork, 
since dependencies can exist between variables of different types.
While we expect correctness, we cannot expect (refutation) completeness, since
 once all desired operations are added the problem becomes undecidable~\cite{DBLP:journals/corr/abs-1711-03363}.

\section{Related Work \& Alternative Approaches}\label{stateoftheart}
In the following, we will briefly present alternative approaches to constraint logic programming.
For a selection of alternative solvers, we will discuss their implementation paradigms, in order to later compare to constraint logic programming.

\subsection{autogen}
autogen~\cite{TAV43} is a specification-based test data generator.
autogen is able to directly use SQL as an input language.
In order to generate test data from it, 
SQL is considered as specification of the SUT and is converted into constraints.
autogen uses an independently developed string constraint solver called CLPQS, 
which handles all requirements stated in \cref{requirements}. 
To support the data types of SQL, autogen interacts with a set of different solvers.
CLPQS represents domains as regular expressions.
One motivation for this paper is to experiment with different representations and propagation algorithms.

\subsection{MiniZinc}
MiniZinc is a solver-independent modeling language 
for constraint satisfaction and optimization problems. 
A MiniZinc model is compiled into a FlatZinc instance 
which can be solved by a multitude of constraint solvers. 
An extension of the MiniZinc modeling language with string variables 
and a set of built-in constraints 
has been suggested by Amadini et al.~\cite{DBLP:journals/corr/AmadiniFPSST16}. 
String variables are defined as words 
over the alphabet of ASCII characters 
and have a fixed, bounded or unbounded length.
Yet, strings are represented as bounded length arrays of integers 
when translating to FlatZinc.
The MiniZinc model itself does allow strings of unbounded length though. 
MiniZinc enables optimization over constraints rather than just satisfiability 
and allows to mix constraints over different domains.
However, there are no direct conversions from other types to strings.

\subsection{SMT Solvers}
SMT solvers such as CVC4~\cite{cvc4} and Z3~\cite{z3} have been used for test case generation in the context of programming languages~\cite{smtfortestcasegeneration}.
Both solvers support constraints over strings and regular expressions 
and are able to handle operations such as concatenation, containment, replacement and constraining the length of strings.

In Z3's original string solver, strings are represented as sequences over bit-vectors. 
The solver itself is incomplete and relies on heuristics. 
In contrast, Z3-str~\cite{z3str} introduces strings as primitive types. 
Z3-str leverages the incremental solving approach of Z3 
and can be combined with boolean and integer constraints. 
There have been several improvements of Z3-str in recent years~\cite{s3,z3str2,z3str3}.

CVC4's string solver~\cite{cvc4str,cvc4str2} allows to mix constraints over strings and integers. 
The authors present a set of algebraic techniques to solve constraints over unbounded strings, usable for arbitrary SMT solvers. 


Another SMT solver for string constraints is \textsc{Trau}~\cite{trau-smt},
which, in contrast to CVC4 and Z3, 
supports context-free membership queries and transducer constraints 
by using pushdown automata. 
\textsc{Trau} implements a Counter-Example Guided Abstraction Refinement (CEGAR) framework, 
 computing over- and under-approximations to improve performance. 
Key idea in \textsc{Trau} is a technique called flattening~\cite{flatten-conquer},
leveraging that (un)satisfiability can be shown using witnesses 
of simple patterns expressable as finite automata. 

\textsc{G-Strings}~\cite{g-strings}
is an extension of the \textsc{Gecode} constraint solver~\cite{schulte2010modeling}. 
Both solvers accept strings of bounded but possibly unknown length. 
In contrast to \textsc{Gecode}, 
strings are not represented using integer arrays 
but as a restricted language of finite regular expressions. 
This prevents the static allocation of possibly large integer arrays 
and, thus, improves performance. 

\subsection{Other Solvers}

Kiezun et al.~presented \textsc{Hampi}~\cite{KiezunGGHE2009}, 
a constraint solver over strings of fixed length 
featuring a set of built-in constraints. 
\textsc{Hampi} is able to reason over regular languages.
String constraints are encoded in bit-vector logic 
which are then solved by the STP~\cite{STPBitvectorSolver} bit-vector solver. 
At the expense of expressiveness, 
limiting the length of strings enables a more restricted encoding,
increasing the performance by several magnitudes. 
However, 
a bit-vector encoding has a larger memory consumption 
than using finite automata. 

Fu et al.\ introduced Simple Linear String Equations (SISE)~\cite{sushi},
a formalism 
for specifying constraints on strings of unbounded length,
and presented the constraint solver \textsc{Sushi}
using finite automata to represent domains.


\subsection{Summary}
In summary, several approaches have been suggested for string constraints.
However, no single approach is able to satisfactorily handle the requirements posed for test data generators described in \cref{requirements}.
A comparison of different solvers considering the features described in \cref{requirements} can be seen in \cref{table:solver_comparison}. 
The requirement of a combined solver states that a direct conversion between strings and other types is provided. 

\begin{table}[t]
\caption{Features of Constraint Solvers. \partialmark\ Indicates Partial Support or Workaround.}%
\label{table:solver_comparison}
\centering
\begin{tabular}{ @{} l c @{\hspace{10px}} c @{\hspace{10px}} c @{\hspace{25px}} c @{\hspace{10px}} c @{\hspace{10px}} c @{} } 
 \toprule
 Solver & \multicolumn{3}{c}{Strings} &  \multicolumn{3}{c}{Combined Solver}  \\
 \cmidrule{2-4}
 \cmidrule{5-7}
  & Unbounded & Unicode & SQL Operations & Integer & Boolean & Real \\
 \midrule
 CLPQS & \cmark & \cmark & \cmark & \cmark & \cmark & \partialmark \\ 
 MiniZinc & \xmark & \xmark & \partialmark & \xmark & \xmark & \xmark \\ 
 CVC4 & \cmark & \xmark & \cmark & \cmark & \xmark & \xmark \\ 
 Z3-str3 & \cmark & \xmark & \partialmark & \cmark & \xmark & \xmark \\ 
 S3 & \cmark & \xmark & \cmark & \cmark & \cmark & \xmark \\ 
 \textsc{Hampi} & \xmark & \xmark & \partialmark & \xmark & \xmark & \xmark \\ 
 \textsc{Sushi} & \cmark & \cmark & \cmark & \xmark & \xmark & \xmark \\ 
 G-\textsc{Strings} & \xmark & \xmark & \partialmark & \xmark & \xmark & \xmark \\ 
 \textsc{Trau} & \cmark & \xmark & \cmark & \cmark & \xmark & \xmark \\
 \bottomrule
\end{tabular}
\end{table}

\section{Constraint Logic Programming Over Strings}%
\label{sec:clpstr}
We implement a constraint logic programming system for strings using Constraint Handling Rules (CHR)~\cite{chrlang} on top of SWI-Prolog~\cite{dblp:journals/corr/abs-1011-5332} 
called \textsc{ConString}.
We use classic constraint propagation to reduce variable domains.
The system supports strings of unbounded length and is coupled with CLP(FD), CLP(R) and CLP(B) to handle integers, reals and booleans respectively.
While not all SQL string operations are implemented yet,
we plan to do so in future.
One goal is to employ different techniques than CLPQS to compare and possibly improve both solvers. 
We think CLP is adequate since there is a multitude of other solvers to build up upon
and since it provides access to all solutions using backtracking.

In the following,
we present our encoding of string domains and discuss its advantages and drawbacks,
followed by
the currently featured constraints, 
selected constraint handling rules and solver integrations. 

\subsection{Domain Definition}\label{domaindefinition}
To fulfill the requirements posed in \cref{requirements},
we decided not to enforce a fixed length of strings and
to use regular expressions
as input.
The employed alphabet consists of
ASCII characters and some special characters
like umlauts and accented characters.
Dynamic character matching is possible
by specifying ranges (e.g., \texttt{[0-9a-f]}),
or by using the dot operator.
We match a whitespace in regular expressions by \texttt{\textbackslash s}
while actual whitespace characters can be used to structure regular expressions
without being part of the accepted language.

Further,
we support the usual regular expression operators on characters,
i.e., quantity operators
(\texttt{*}, \texttt{+} and \texttt{?})
and the alternative choice operator (\texttt{|}).
For convenience,
our regular expressions offer more strict repetition definitions noted by
\texttt{\{n\}} (exactly \texttt{n} times),
\texttt{\{m,n\}} (\texttt{m} to \texttt{n} times) and
\texttt{\{m,+\}} (at least \texttt{m} times).

\subsection{Domain Representation}\label{domainrepresentation}
Since \textsc{ConString} is supposed to handle strings of unbounded length, 
we represent domains as finite automata as done by Golden et al.~\cite{CROverStrings}. 
First, 
this allows for a concise specification of regular languages with low memory consumption. 
Second, 
finite automata support basic operations 
such as union, intersection, concatenation or iteration 
and are closed under each of these operations. 
In particular, 
we use non-deterministic finite automata 
with $\epsilon$-transitions. 

Since SWI-Prolog does not have a native library for handling finite automata, we encode them as a self-contained term \texttt{automaton\_dom/4} 
consisting of 
a set of states, 
a transition relation 
as well as a set of initial and final states. 
The states are a coherent list of integers $1\dots n$, $n\in\mathbb{N}$. 
The transition relation is implemented as a list of triples 
containing a state \(s_1\), 
a range of characters (might contain a single character only) 
and a target state \(s_2\) reached after processing a character
from the range of characters in \(s_1\), 
e.g., \texttt{(0, a, 1)}. 

We implement the common operations on finite automata used for regular languages 
as well as basic uninformed search algorithms 
used to label automata, 
i.e., to find a word having an accepting run. 
The search is backtrackable 
providing access to an automaton's complete language.

\subsubsection{Efficiency}%
\label{sec:efficiency}
The chosen representation of finite automata has several drawbacks. 
We use lists to store states and transitions 
providing linear time  
concatenation and element access 
leading to a loss of performance, 
especially when labeling automata.
It would be desirable to use a data structure such as hashsets,
which provide amortized constant time performance for basic operations.
However, such a data structure is currently unavailable in SWI-Prolog\footnote{While SWI-Prolog has built-in support for dictionaries, 
element access is logarithmic and updates are linear in size.}.

Another drawback is 
that we have to rename states 
when performing basic operations on automata.
For instance, 
the concatenation $\mathcal{A}_{1}.\mathcal{A}_{2}$ is implemented 
by using the final states of $\mathcal{A}_{2}$ for the resulting automaton 
and adding an $\epsilon$-transition from all final states of $\mathcal{A}_{1}$ 
to all initial states of $\mathcal{A}_{2}$.
In order to avoid ambiguities, the states of $\mathcal{A}_{2}$ have to be renamed 
by shifting their identifier names by the number of states in \(\mathcal{A}_1\).
This renaming is one of the main issues for efficiency
as it adds a linear time complexity component
with respect to the size of the second automaton
to all of the basic operations.

\subsection{Constraint Handling Rules}\label{string_chr}
We use CHR on top of SWI-Prolog 
providing the constraint store 
and propagation unit to reduce variable domains. 
Moreover, CHR serves as user interface. 

The CHR language is committed-choice, 
i.e., once a rule is applied it cannot be revoked by backtracking. 
Rules consist of three parts: 
a head, a guard and a body. 
A rule is triggered as soon as the head matches constraints in the constraint store.
Guards allow to impose restrictions on rule execution.
Finally,
the body consists of Prolog predicates 
and CHR constraints. 
Predicates are called as usual
while constraints are added to the constraint store,
possibly triggering further propagation. 
All available constraints are propagated 
until the constraint store reaches a fix point.
Solving fails 
if an empty string domain is discovered. 

CHR provides three different kinds of rules: 
First, propagation rules of the form 
\texttt{head ==> guard | body},
where the body is called if the guard is true.
The head constraints are kept.
Simplification rules of the form 
\texttt{head <=> guard | body} 
update the constraint store by
replacing the head constraints
by those derived from the body. 
Simpagation rules of the form 
\texttt{head1 \textbackslash{} head2 <=> guard | body} 
are combined rule, retaining some constraints of the head.

Our implementation currently supports 
several basic operations on regular languages like intersection, concatenation or iteration 
as well as a membership constraint, 
arithmetic length constraints (fixed or upper bound), 
string to integer conversion, 
prefix, suffix and infix constraints 
and case sensitivity constraints.
For now, we ensure arc- and path-consistency of our constraints.
In the following, 
we will describe selected constraint handling rules in more detail. 

The membership constraint 
is defined as shown in~\cref{chr:str_in}. 
The first rule is applied in case membership is called 
with a string or regular expression. 
Then, 
a finite automaton representing the input domain is generated 
and the same constraint is applied to this automaton domain. 
The second rule states
that whenever a domain is empty 
constraint solving should fail as no solution exists.
Third, 
in case the string domain becomes constant, we propagate the value to the variable. 
The fourth rule joins two non-equal membership constraints for the same variable 
by intersecting both domains 
and replacing the two constraints 
by a single membership constraint. 
A final rule is used to remove one of two identical membership constraints.
Note 
that 
\texttt{gen\_dom/2} and \texttt{intersection/3} 
are called for internal domain computation
and not added to the constraint store.


\begin{lstlisting}[float=t, label=lst:str_in_rules, label=chr:str_in, caption=CHR rules for the membership constraint \texttt{str\_in/2}.]
str_in(S1, S2) <=>
    string(S2) | gen_dom(S2, D),str_in(S1, D).
str_in(_, D) ==> is_empty(D) | fail.
str_in(Var,D) ==> D = string_dom(Cst) | Var = Cst.
str_in(S, D1), str_in(S, D2) <=>
    D1 \= D2 | intersection(D1, D2, D3), str_in(S, D3).
str_in(S, D1)\ str_in(S, D2) <=>  D1 == D2 | true.
\end{lstlisting} 

Concatenation is defined using two rules 
as shown in \cref{chr:str_concat}.
It relies on the membership constraint 
by assuming that two \texttt{str\_in/2}
refer to different variables. 
The first rule defines the concatenation 
of two different string variables 
by concatenating their automata domains 
and adding a new membership constraint for the result. 
Analogously,
the second rule defines the concatenation of the same string variable onto itself. 
In order to efficiently propagate a constant string result to the first two arguments, 
we add a third rule using SWI-Prolog's string concatenation, e.g., \texttt{string\_concat(A, B, "test")}, 
providing all solutions on backtracking. 
If a candidate has been found, 
it is checked upon labeling 
whether the candidate is accepted by the corresponding domains. 
If so, 
membership constraints are propagated 
assigning constant values to all arguments. 

\begin{lstlisting}[float=t, label=lst:str_in_rules, label=chr:str_concat, caption=CHR rules for the concatenation constraint \texttt{str\_concat/3}.]
str_in(S1, D1), str_in(S2, D2), str_concat(S1, S2, S3) ==>
    concat(D1, D2, D3), str_in(S3, D3).
str_in(S1, D1), str_concat(S1, S1, S3) ==>
    concat(D1, D1, D3), str_in(S3, D3).
\end{lstlisting}

The iteration operation \texttt{str\_repeat/[2,3,4]} 
is defined as repeated concatenation. 
Case sensitivity operations 
are defined by setting up membership constraints 
to generated domains accepting only upper or lower case characters. 

The infix operation \texttt{str\_infix/2} 
for two string variables $s_1$ and $s_2$ 
is defined by adding a membership constraint for $s_1$ to be an element of the regular language $\mathcal{L}(.^*).\mathcal{L}(s_2).\mathcal{L}(.^*)$ as shown in \cref{chr:str_infix}.
Again, the first rule is a wrapper generating a finite automaton domain 
from a string or regular expression. 
Prefix and suffix operations are defined in the same manner.
 
\begin{lstlisting}[float=t, label=lst:str_in_rules, label=chr:str_infix, caption=CHR rules for the infix constraint \texttt{str\_infix/2}.]
str_infix(S, IStr) <=>
    string(IStr) | gen_dom(IStr, IDom), str_infix(S, IDom).
str_infix(S, IDom) <=>
    any_char_dom(A), repeat(A, AStar),
    concat(IDom, AStar, T), concat(AStar, T, ResDom),
    str_in(S, ResDom).
\end{lstlisting}

\subsection{Integration of CLP(FD), CLP(R) and CLP(B)}
In order to enable the generation of richer test data 
and allow for a greater coverage of test scenarios, 
we extend \textsc{ConString} 
to support combining constraints over different domains. 
In particular, we support constraints over finite domain integers 
using CLP(FD)~\cite{clpfd_swi}, 
constraints over reals using CLP(R) 
and constraints over booleans using CLP(B)~\cite{clpb,clpb2}. 
As an interface, we provide the bidirectional constraints \texttt{str\_to\_int/2}, \texttt{str\_to\_real/2} and \texttt{str\_to\_bool/2}.

The implementation of \texttt{str\_to\_int/2} consists of four rules 
as shown in \cref{chr:str_to_int_basic}. 
In order to detect failure early
we check for inequality 
if both arguments are constants. 
If only the integer variable is a constant, 
we convert and assign the value to the string. 
In the third rule, 
a constant string is assigned to the integer variable. 
Note that \texttt{number\_string/2} removes leading zeros by default.
Besides that, we provide a rule 
to fail for constant strings not representing integers.

We additionally provide a second implementation \texttt{str\_to\_intl/2} 
allowing leading zeros 
in order for constraints such as \texttt{str\_to\_int("00", 0)} to hold. 
This is achieved by concatenating the domain of \texttt{$0^{*}$} to \texttt{IDom} in line 6 of \cref{chr:str_to_int_basic}.

\begin{lstlisting}[float=t, label=chr:str_to_int_basic, caption=Basic rules for the integration of CLP(FD) propagating constant values.]
str_to_int(S,I) ==>
  string(S), integer(I), number_string(SInt, S), I \== SInt|
  fail.
str_to_int(S,I) ==>
  integer(I), number_string(I, IString)|
  cst_str_dom(IString, IDom), str_in(S, IDom).
str_to_int(S,I), str_in(S,D) ==>
  D = string_dom(CstString),
  number_string(CstInteger, CstString)| I #= CstInteger.
str_to_int(S,_), str_in(S,D) ==>
  D = string_dom(CstString), \+ number_string(_, CstString)|
  fail.
\end{lstlisting}

The integration of CLP(R) and CLP(B) is implemented analogously 
propagating membership constraints to a specific backend 
if variables are constant values. 
Again, 
alternative implementations are provided 
allowing an arbitrary amount of leading zeros 
when converting from string to boolean or real.

\section{Case Studies}%
\subsection{Generation of IBAN Numbers}
\label{sec:iban}
As a case study, 
we specify the computation of valid International Bank Account Numbers (IBANs) 
as a constraint system as done by Friske and Ehmke~\cite{TAV43}. 
This example is of interest as it yields a relatively large search space 
and requires the conversion between integer and string. 
Generated data can, for instance, be used to initialize unit tests of components validating IBANs.

A german IBAN consists of 22 characters which are characterized as follows:
The first two characters represent the country code (here, the constant ``DE'')  
while the third and fourth characters are a checksum. 
The remaining 18 digits represent the Basic Bank Account Number (BBAN). 

We can compute valid IBANs using a given country code as follows: 
Represent the country code as a digit where ``A'' equals 10, ``B'' equals 11, etc.
The german country code ``DE'' is hence encoded as 1314. 
Concatenate two zeros to the encoded country code (i.e., 131400) 
and prepend the BBAN. 
This forms a 24 digit number, \(\sigma_b\).
In order to compute the valid checksum $\sigma_c$, 
the constraint $98 - (\sigma_b ~mod~ 97) = \sigma_c$ must hold.
Finding a solution binds the BBAN to a value in its domain
and provides its corresponding checksum \(\sigma_c\). 
To derive the actual BBAN,
remove the suffix ``131400''.
Finally, 
concatenate the computed checksum \(\sigma_c\) with the BBAN 
and prepend the country code ``DE'' as a string. 

The complete constraint system is shown in~\cref{example:iban_clpstr}. 
Lines 3 and 4 define the BBAN and the 24 digits number $\sigma_b$ respectively. 
The constraint for computing \(\sigma_c\) is set in line 5.
The remaining specification is straightforward as described above. 
Note that we allow leading zeros for the checksum's string.

\begin{lstlisting}[float=t, label=lst:str_in_rules, label=example:iban_clpstr, caption=Constraint system to compute all valid german IBANs.]
iban(IBAN) :-
    SigmaC in 0..96,
    BBAN in 100000000000000000..999999999999999999,
    SigmaB #= BBAN * 1000000 + 131400, 
    SigmaB mod 97 #= SigmaC,
    str_label([SigmaB, SigmaC]),
    str_to_int(BBANStr, BBAN),
    CheckSum #= 98 - SigmaC,
    str_to_intl(CheckSumStr, CheckSum),
    str_size(CheckSumStr, 2),
    str_in(DE, "DE"),
    str_concatenation(DE, CheckSumStr, IBANPrefix),
    str_concatenation(IBANPrefix, BBANStr, IBAN),
    str_label([IBAN]).
\end{lstlisting}

To benchmark we generate sets of IBANs of varying sizes,
using an Intel Core i7-6700K with 16GiB RAM. 
We used SWI-Prolog's predicate \texttt{statistics/2} to measure the walltime. 
\cref{iban_benchmarks} shows the median time of five independent runs
and compares our solver with CLPQS.
As can be seen, \constring{} performs overall slightly better than CLPQS 
with the exception of the generation of 250,000 IBANs.
Up to one thousand samples both solvers appear to scale linearly.
Notable exception is the jump from 10,000 to 100,000 generated samples.
Here, both solvers scale worse: CLPQS scales with a factor of 47,
whereas \constring{} takes 50 times as long as for generating 
10,000 IBANs instead of the expected factor of 10.
At least for \constring{},
experimental results have shown 
that this non-linear growth
is caused by SWI-Prolog's CLP(FD) library. 

\begin{table}[t]
\caption{Benchmarks for generating IBANs. Walltime in seconds.}%
\label{iban_benchmarks}
\centering
\begin{tabular}{ c c c c c c c c } 
  \toprule
  Amount       &     1 &    10 &   100 &  1,000 &  10,000 &  100,000 & 250,000 \\
  \midrule
  CLPQS       & 0.006 & 0.024 & 0.240 & 2.029 & 32.163 & 1525.457 & 9261.204 \\
  \constring{} & 0.007 & 0.038 & 0.105 &  1.066 &  26.573 & 1342.597 & 9841.225 \\
  \bottomrule
\end{tabular}
\end{table}

We also encoded the example in SMT-LIB 
to compare \constring{} and CLPQS with Z3-str3 and CVC4. 
Unfortunately, CVC4 did not return a result but timed out after 600 seconds.
Z3-str3 found a single solution in around 0.2 seconds.
We were unable to compute multiple solutions using Z3-str3.

\subsection{Generating of Calender Dates}%
\label{sec:calender-dates}

Another example is the generation of various date expressions, 
which is of interest for testing for many tools
which need to parse valid dates and reject invalid ones.
The accepted expressions are of either of the forms
\enquote{Tuesday},
\enquote{August 30},
\enquote{Tuesday, August 30},
\enquote{August 30, 2016}, or
\enquote{Tuesday, August 30, 2016}.

\cref{example:dates} shows the corresponding constraints, taken
from~\cite[Section 3]{karttunen1996regular}.
The CSP consists of defining the basic building blocks first:
the weekdays, the months, and valid year numbers.
Thus, only the years 1 to 9999 are accepted.
Further, the more complex parts are constructed,
each consisting of a combination of operations on variables constrained before.
This leads up to the final definition of \texttt{Date},
which is a union of all possible notations described above.

\begin{lstlisting}[float=t, label=example:dates, caption=Constraint system to compute diverse calendar date expressions]
date(Date) :-
  WeekDay str_in "Monday|Tuesday|...|Sunday",
  Month str_in "January|February|...|December",
  Day str_in "[1-9]|[1-2][0-9]|3[0-1]",
  Year str_in "[1-9][0-9]{0,3}",
  MonthDay match Month + "_" + Day,
  MonthDayYear match MonthDay + ",_" + Year \/ MonthDay,
  FullDate match WeekDay + ",_" + MonthDayYear,
  Date match MonthDayYear \/ FullDate \/ WeekDay,
  str_label([Date]).
\end{lstlisting}

Note that we employ a shorthand notation for the setup of constraints.
\texttt{MonthDayYear} for example has to match the language defined by the union of
the \texttt{MonthDay} domain and
the concatenation of \texttt{MondDay}, a separator, and the \texttt{Year}.
This shorthand notation allows for a more readable definition of CSPs.

\begin{table}[htb]
\caption{Benchmarks for generating date expressions. Walltime in seconds.}
\label{table:date-benchmarks}
\centering
\begin{tabular}{ c c c c c c c } 
  \toprule
  Amount       &     1 &      10  &    100 &   1,000 &  10,000 &  100,000 \\
  \midrule
  CLPQS        & 	 0.000  & 0.000 &   0.000 &  0.000 &   0.000 &     0.010 \\
  \constring{} &  0.010   &  0.010 &  0.010 &  0.011 &  0.080  &   0.965   \\
  \bottomrule
\end{tabular}
\end{table}

\cref{table:date-benchmarks} shows a brief performance evaluation as done in \cref{sec:iban}. 
As can be seen, CLPQS is notably faster than \constring{}. 
The automata created by \constring{} are probably large 
due to the alternative choice operator and the union operator 
leading to a lack of performance when labeling data.
Reducing the size of automata, 
e.g., by removing $\epsilon$-transitions, 
will likely increase performance.

We also encoded the example in SMT-LIB to compare \constring{} and CLPQS with Z3-str3 and CVC4.
Z3 found a single solution in around 0.070 seconds 
while CVC4 took around 0.084 seconds. 
We were unable to compute multiple solutions with both solvers.

\section{Way Forward \& Future Work}%
\label{sec:conclusion}
\subsection{An Efficient Backend}
For classic domain propagation to work on strings, an efficient representation of possible values is needed.
So far, we represent automata as outlined in \cref{domainrepresentation}.
As discussed, this is not the most efficient approach, as
certain algorithms need to traverse the list of states or transitions to find a particular one.

Other known automaton libraries 
such as dk.brics.automaton~\cite{automaton} feature more efficient representations and algorithms.
However, these are usually based on using pointers or objects and cannot easily be ported to Prolog for obvious reasons.
At the same time, connecting the Java or C ports of the library to our Prolog system leads to all kinds of difficulties when it comes to proper handling of backtracking. 
Moreover, Prolog programs are no longer declarative when using stateful data structures 
without cloning data after each operation.

In future, we want to experiment with porting dk.brics.automaton or a comparable library to Prolog while retaining its efficiency.
So far, we have different approaches in mind.
First, we could implement low-level data structures outside of Prolog (e.g.,~in C) and render them backtrackable using a thin Prolog layer.
Second, we could mimic the internal workings of the library using, i.e., attributed variables to store (mutable) class variables and links to other \enquote{objects}.
  While this would avoid possible backtracking issues, it would not be as idiomatic.

Furthermore, 
we want to evaluate whether it is more efficient to use deterministic finite automata 
or, in general, $\epsilon$-free automata.

Domain representation aside, so far we did not provide options for labeling, e.g.,
concerning the enumeration order.
Additional options like 
enumerating a string domain in alphabetical, 
reversed alphabetical 
or a randomized order  
will most likely improve performance 
for some constraint satisfaction problems. 
This would also enable to provide different distributions of test data for a given domain.
Yet, labeling options of integrated solvers like CLP(FD) can be used.

\subsection{Combining Solvers}
\label{subsection:combining_solvers}
Of course, a solver like the one we outlined above would still be too weak to efficiently support the constraints we discussed in \cref{introduction}.
In consequence, we envision an integration of a CLP-based solver and the other solvers discussed in \cref{stateoftheart} into a combined solving procedure.
This could be done following the approach we used for first-order logic in~\cite{probz3integration}.

A more simple strategy would be to use multiple solvers at once and returning the first result computed.
This will have a performance benefit, given that the solvers described in \cref{stateoftheart} 
have diverse approaches and mixed performances in certain situations.
Implementing such a portfolio is somewhat complicated, since there is no standardized interface for constraint solvers~\cite{cvc4str2}, leading to a large overhead translating constraints in between solvers.
However, 
a promising draft for an interface~\cite{SMTLibStrings}
has been proposed recently.

\subsection{Further Studies}

In this paper we showed two small and realistic use cases.
We did not show a complex case study like an open interval concatenation.
In future work we will address more involved case studies to examine the limits of string constraint solving.

\section{Conclusion}
In this paper, we discussed how synthetic test data can be generated and what the common pitfalls are.
We discussed currently available solvers over strings and outlined that string constraint solving has made considerable progress recently.
However, hurdles remain and generation of artificial test data remains complicated at least.

We implemented an, albeit simple, prototype of a string constraint solver based on constraint logic programming and classical domain propagation.
While it does not yet offer all features desired, our prototype shows that our approach is feasible and promising.

However, we believe that no single solver will be able to handle all requirements sufficiently and that reimplementing features commonly found in other solvers might not be worthwhile.
In consequence, we think that an integration of solvers such as the one discussed in \cref{subsection:combining_solvers} is very promising and we hope to be able to lift our results for first-order-logic to string domains in the future.

\bibliographystyle{abbrv}
\bibliography{paper}

\end{document}